\documentclass[preprint,prl]{revtex4}%%%% change preprint to twocolumn ... but for submission must be preprint
\usepackage[latin1]{inputenc}

\usepackage{amssymb}
\usepackage{graphicx}
\usepackage{epsfig}
\usepackage{psfrag}
\usepackage{latexsym}
\usepackage{indentfirst}
\usepackage{fancyhdr}
\usepackage{amssymb}
\usepackage{amsmath}
\usepackage{amsfonts}
\usepackage{colordvi}
\usepackage{pifont}
\usepackage{color}
\usepackage{graphicx}
\usepackage{url}
\usepackage{array}
\usepackage{rotating}

\def\beq{\begin{equation}}
\def\eeq{\end{equation}}
\def\be{\begin{equation}}
\def\ee{\end{equation}}
\def\bea{\begin{eqnarray}}
\def\eea{\end{eqnarray}}

\begin{document}
\title{Lepton Mixing Predictions from $\Delta(6n^2)$ Family Symmetry}

\author{Stephen F. King}
\email{S.F.King@soton.ac.uk}
\affiliation{School of Physics and Astronomy, \\ University of Southampton, Southampton, SO17 1BJ, U.K.}

\author{Thomas Neder}
\email{T.Neder@soton.ac.uk}
\affiliation{School of Physics and Astronomy, \\ University of Southampton, Southampton, SO17 1BJ, U.K.}

\author{Alexander J. Stuart}
\email{A.Stuart@soton.ac.uk}
\affiliation{School of Physics and Astronomy, \\ University of Southampton, Southampton, SO17 1BJ, U.K.}

\begin{abstract}
We obtain predictions of lepton mixing parameters for direct models based on 
$\Delta(6n^2)$ family symmetry groups for arbitrarily large $n$
in which the full Klein symmetry is identified as a subgroup
of the family symmetry.  After reviewing and developing the group theory associated with $\Delta(6n^2)$, we find many new candidates for large $n$ able to yield 
reactor angle predictions within $3\sigma$ of recent global fits. We show that such $\Delta(6n^2)$ 
models with Majorana neutrinos predict
trimaximal mixing with reactor angle $\theta_{13}$ fixed up to a discrete choice, 
an oscillation phase of either zero or $\pi$ and the 
atmospheric angle sum rules 
$\theta_{23}=45^\circ \mp \theta_{13}/\sqrt{2}$, respectively, which are consistent with recent global fits 
and will be tested in the near future.
\end{abstract}

\maketitle

%%%%%%%%%%%%%%%%%%%%%%%%%%%%%%%%%%%

\section{Introduction}

%%%%%%%%%%%%%%%%%%%%%%%%%%%%%%%%%%%

The measurement of a rather large reactor mixing angle by the Daya Bay~\cite{DayaBay}, RENO~\cite{RENO}, and Double Chooz~\cite{DCt13} collaborations adds further complexity to an already difficult puzzle of flavour. Perhaps the best way to address this dilemma is to utilise the methods developed in the era of an unmeasured reactor angle and introduce an additional discrete family symmetry, $\mathcal{G}_f$, under which all fields transform. This family symmetry will then be spontaneously broken in order to generate the observed fermionic masses and mixings \cite{pdg}.   However before even considering the construction of a model, it may be insightful to know some of the possible candidate symmetries for $\mathcal{G}_f$.  Herein lies the goal of this work, shedding light on a particular class of  candidates for $\mathcal{G}_f$, i.e. the $\Delta(6n^2)$ groups.

In the following text, we demand that the discrete group $\mathcal{G}_f$ be a subgroup of the continuous group $SU(3)$ (or $U(3)$) because its fundamental representation is $3$-dimensional.  We further restrict ourselves to working with the $\Delta(6n^2)\cong (Z_n\times Z_n)\rtimes S_3$ subgroups of $SU(3)$ due to the past and current popularity of $S_4\cong\Delta(24)$ ($n=2$) in flavour model building  (see \cite{King:2013eh} and references contained therein) as well as recent publications demonstrating that $\Delta(96)$ ($n=4$)\cite{d96}, $\Delta(150)$ ($n=5$)\cite{d150lam,d150lam2}, $\Delta(600)$ ($n=10$)\cite{d150lam2,Holthausen:2012wt} and $\Delta(1536)$ ($n=16$)\cite{Holthausen:2012wt}
generate phenomenologically viable predictions for the lepton mixing angles.   

In contrast to the above studies, here we consider the infinite series of groups $\Delta(6n^2)$ as candidates for a family symmetry group. We exploit the presentation of the $\Delta(6n^2)$ series to elucidate all possible mixing patterns resulting from all possible Klein subgroups for each $\Delta(6n^2)$ group where $n$ is an even integer.
We consider only $\Delta(6n^2)$ groups where $n$ is even because these are the only cases which can contain a \textit{complete} Klein subgroup, i.e. all \textit{four} Klein subgroup elements with generators denoted by 
$S,U$, where the invariance of the
neutrino mass matrix under the Klein symmetry group is sufficient to completely 
fix the neutrino mass matrix for Majorana neutrinos
(for a review see e.g.  \cite{King:2013eh}).  
We remark that $\Delta(6n^2)$ family symmetry models in which the Klein
symmetry is not identified as a subgroup of the family symmetry have also been studied 
(see e.g. \cite{King:2009ap}).

Thus with the preliminary assumptions and goals of this work put forth, we proceed by introducing the framework in which we will work.  Afterwards, a brief review of the representations of $\Delta(6n^2)$ will be given.  Finally, the details of our method elucidated and the results presented.

%%%%%%%%%%%%%%%%%%%%%%%%%%%%%%%%%%%%%%%%%%%%%%%%%%%%%%%%%%%%

\section{From $\mathcal{G}_f$ to Lepton Mixing}

%%%%%%%%%%%%%%%%%%%%%%%%%%%%%%%%%%%%%%%%%%%%%%%%%%%%%%%%%%%%

As previously mentioned, to address the puzzling issue of flavour, we will introduce a discrete family symmetry which will be spontaneously broken to different subgroups in the charged lepton and neutrino sectors, thereby generating the observed lepton masses and mixings.  In such a \textit{direct model} of flavour \cite{King:2013eh}, the family group  is broken to some abelian subgroup $Z_m^T$ ($m$ an integer) in the charged lepton sector and to the $Z_2^S\times Z_2^U$ Klein Symmetry Group in the neutrino sector.  The superscripts denote that $S$, $T$ and $U$ are the generators of their corresponding $Z_m$ group in the diagonal charged lepton basis. Hence, the   
$Z_2^S\times Z_2^U$ transformations on $\nu_L$ and the $Z_m^T$ transformations on $e_{L,R}$
leave the Lagrangian invariant.
This implies that 
\begin{equation}
[S,M^\nu]=[U,M^\nu]=0 \text{ and } [T,M^e]=0,
\label{GcommuteM}
\end{equation}
where $M^{\nu}$ and $M^e$ represent the mass matrices multiplied by their Hermitian conjugates.
Since $S$ and $U$  commute with $M^{\nu}$  they are diagonalised by the same matrix $V^{\nu}$.
Similarly $T$ and  $M^e$ are diagonalised by the same matrix $V^{e}$.
Since $M^{\nu}$ and $M^e$ relate to the left-handed fields, the PMNS matrix is then given by
\begin{equation}
V=V^{e\dagger}V^\nu \label{VPMNS}.
\end{equation}
To obtain the matrices $V^{\nu}$ and $V^{e}$, and hence the PMNS matrix, we only need to identify the
generators $S$, $U$ and $T$ and diagonalise them. In practice, this amounts to finding the eigenvectors 
of $S$, $U$ and $T$ which form the columns of $V^{\nu}$ and $V^{e}$. 
This is straightforward for $T$ since the eigenvalues are non-degenerate due to the fact that 
$T$ must be an element of $\mathcal{G}_f$ of order 3 or greater.
However for the $S$ and $U$ generators the situation is slightly different because they are  $3\times 3$ matrices of \textit{order 2}.  Thus, each eigenvalue of $S$ or $U$ can only be $\pm 1$. Without loss of generality, we choose $\det(S)=\det(U)=+1$, so that each generator has two $-1$ eigenvalues, rendering the corresponding eigenvectors non-unique. Since the three matrices $S$, $U$ and $SU$ each have one (unique) $+1$ eigenvalue this allows for the calculation of three unique eigenvectors (one for each non-trivial Klein group generator), each providing an
$i$th column of the matrix $V^\nu$:
 \begin{equation}
G_i V^\nu_i=+ V^\nu_i\text{, for }G_i \in \{S,U,SU\}.
\label{columns}
\end{equation}
In this way all three columns of $V^{\nu}$ can be obtained.

The remarkable method outlined in this section enables the calculation of the lepton mixing matrix by only considering the family group's representation matrices \cite{d150lam,Holthausen:2012wt}. However, this certainly requires explicit representation matrices for the $\Delta(6n^2)$ group's representations.  We construct these in the next section.

%%%%%%%%%%%%%%%%%%%%%%%%%%%%%%%%%%%%%%%%%%%%%%%%%%%%

\section{The Group Theory of $\Delta(6n^2)$}

%%%%%%%%%%%%%%%%%%%%%%%%%%%%%%%%%%%%%%%%%%%%%%%%%%%%

The  $\Delta(6n^2)$ groups are finite non-Abelian subgroups of $SU(3)$ ($U(3)$)  of order $6n^2$.  They are isomorphic to the semidirect product~\cite{Escobar:2008vc},
\begin{equation}\label{semidirect}
\Delta(6n^2)\cong (Z_n\times Z_{n})\rtimes S_3.
\end{equation}
The Klein group $Z^S_2\times Z^U_{2}$ (in direct models) can either originate purely from the 
$Z_n\times Z_{n}$ or it will involve the $S_3$ generators as well, both possibilities requiring even $n$.
We may re-express Eq. (\ref{semidirect}) in a more illuminating form by taking advantage of the structure of $S_3$:
\begin{equation}\label{semidirectredef}
\Delta(6n^2)\cong  (Z_n^c\times Z_{n}^d) \rtimes(Z_3^a\rtimes Z_2 ^b) .
\end{equation}
Notice that in Eq. (\ref{semidirectredef}), $(Z^c_n\times Z^d_{n})$ forms a normal, abelian subgroup of $\Delta(6n^2)$, generated by the elements $c$ and $d$, and  $(Z_3^a\rtimes Z_2 ^b)$ is nothing more than $S_3$ rewritten in terms of its generators $a$ and $b$. From Eq. (\ref{semidirectredef}) follows that the relevant presentation of $\Delta(6n^2)$ is~\cite{Escobar:2008vc}:
\begin{equation}
\begin{split}
a^3=b^2=(ab)^2&=c^{n}=d^{n}=1, \ \ cd=dc,\\\
aca^{-1}=c^{-1}d^{-1}&,\hspace{5pt} ada^{-1}=c,\\
bcb^{-1}=d^{-1}&,\hspace{5pt} bdb^{-1}=c^{-1}.
\end{split}\label{d96pres}
\end{equation}

Another advantage of the presentation in Eqs. (\ref{semidirect})-(\ref{semidirectredef}) is that every group element can be written as 
\begin{equation}
g=a^{\alpha}b^{\beta}c^{\gamma}d^{\delta},
\label{elementsstandardorder}
\end{equation}
with $\alpha=0,1,2$, $\beta=0,1$ and $\gamma,\delta=0,\ldots,n-1$, making the computation of all group elements for a certain representation/basis computationally simple.  All that needs to be known next is the explicit forms of generators.

In order to find the explicit forms for the generators, we restrict ourselves to 3-dimensional irreducible representations of $\Delta(6n^2)$. Then, it can be shown that $\Delta(6n^2)$ has $2(n-1)$ 3-dimensional irreducible representations denoted by ${\bf 3}_k^l$ and explicitly generated by \cite{Escobar:2008vc}:
\begin{eqnarray}
\begin{array}{cc}
a=\begin{pmatrix}0&1&0\\0&0&1\\1&0&0\end{pmatrix},& b=(-1)^{k+1}\begin{pmatrix}0&0&1\\0&1&0\\1&0&0\end{pmatrix},\\&\\c=\begin{pmatrix}\eta^l&0&0\\0&\eta^{-l}&0\\0&0&1\end{pmatrix},&d=\begin{pmatrix}1&0&0\\0&\eta^l&0\\0&0&\eta^{-l}\end{pmatrix},
\end{array}\label{explicitd96}
\end{eqnarray}
where $\eta=e^{2\pi i/n}$; $k=1,2$; and $l=1,\ldots, n-1$. 

We further restrict ourselves to working with only faithful irreducible representations of $\Delta(6n^2)$.  Thus, we exclude all representations in Eq. (\ref{explicitd96}) where $l$ divides $n$, as they are unfaithful. Of the remaining representations,  ${\bf 3}_k^l$ and ${\bf 3}_k^{l{^\prime}}$ are complex conjugates of each other if $l+l^\prime=n$.  Therefore, they will provide complex conjugated mixing matrices. 
The remaining representations provide the same sets of mixing matrices  because the generators $a$ and $b$ are the same for all $l$ and
\begin{equation}\label{cdreprel}
c({\bf 3}_k^l)=c({\bf 3}_k^1)^l \text{ and }d({\bf 3}_k^l)=d({\bf 3}_k^1)^l.
\end{equation}
Then, by considering Eq. (\ref{elementsstandardorder}) and Eq. (\ref{cdreprel}) it is clear that each power of the $c$ and $d$ generators will appear in every 3-dimensional irreducible representation.  For these reasons, it suffices if one only considers $S$, $T$, and $U$ as representation matrices from ${\bf 3}_2^1$. Notice that $k=2$ has been chosen because in this case the determinant of the elements of order 2 is +1. 

Having reduced the possible cases needed for consideration, the next step is to calculate all Klein subgroups of $\Delta(6n^2)$.  This is accomplished by first calculating all order two elements.  From the generators and rules given in Eq. (\ref{d96pres}) it follows that all order 2 elements in $\Delta(6n^2)$ are given by:
\begin{equation}\label{order2elem}
c^{n/2},~d^{n/2},~c^{n/2}d^{n/2},~bc^\epsilon d^\epsilon,~abc^\gamma,~\text{and}~a^2bd^\delta,
\end{equation}
where $\epsilon,\gamma, \delta=0,\dots, n-1$.

The order 2 elements found in Eq. (\ref{order2elem}) serve as a starting point for calculating Klein Symmetry groups of $\Delta(6n^2)$.  Using Eq. (\ref{d96pres}) and Eq. (\ref{order2elem}) reveals the Klein subgroups 
of $\Delta(6n^2)$ for even $n$ to be:
\begin{eqnarray}\label{numklein1}
\{1,c^{n/2},d^{n/2},c^{n/2}d^{n/2}\},\\\label{numklein2} \{1,c^{n/2},abc^{\gamma'},abc^{\gamma'+n/2}\},\\\label{numklein3}
 \{1,d^{n/2},a^2bd^{\delta'},a^2bd^{\delta'+n/2}\}, \\\label{numklein4} \{1,c^{n/2}d^{n/2},bc^{\epsilon'} d^{\epsilon'}, bc^{\epsilon'-n/2}d^{\epsilon'-n/2}\},
\end{eqnarray}
where $\gamma', \delta', \epsilon'=1,\ldots, n/2$. Notice that Eq. (\ref{numklein1}) corresponds to the Klein symmetry
originating completely from $Z_n\times Z_{n}$ whereas Eqs. (\ref{numklein2})-(\ref{numklein4}) involve also $S_3$.
In the basis of Eq. (\ref{explicitd96}), one of the Klein generators (taken to be $S$) is diagonal for all
cases, while in the case of Eq. (\ref{numklein1}) both Klein generators $S,U$ are diagonal \footnote{As an example of the Klein subgroups in Eqs. (\ref{numklein1})-(\ref{numklein4}), in $\Delta(96) ($n=4$) $\cite{d96}, it was found that 
for the bi-trimaximal mixing example $S=d^2$ and $U=a^2bd^3$, implying that these generators are contained in the Klein subgroups defined in Eq. (\ref{numklein3}).}. 

As previously discussed, the $T$ generator which controls the charged lepton sector must 
be at least of order 3. As shown in the Appendix, only the minimal order 3 case is phenomenologically viable
and so we only consider this possibility. 
In $\Delta(6n^2)$ groups where 3 does not divide $n$, all elements of order 3 are expressible as \cite{Escobar:2008vc}:

\begin{equation}\label{Tcan1}
a c^\gamma d^\delta, a^2 c^\gamma d^\delta
\end{equation}
where $\delta,\gamma=0\dots n-1$  
\footnote{When $n$ is divisible by $3$, there exist more order three elements given by
$c^{n/3}$, $c^{2n/3}$, $d^{n/3}$, $d^{2n/3}$, $c^{n/3}d^{n/3}$, $c^{2n/3}d^{n/3}$, $c^{n/3}d^{2n/3}$,
$c^{2n/3}d^{2n/3}$.
In the basis of Eq. (\ref{explicitd96}), 
these are diagonal matrices of phases. Since $S$ is also diagonal in this basis,  
this would result in phenomenologically unacceptable predictions for leptonic mixing.}.  
Without loss of generality we may choose the order three generator to be 
\begin{equation}\label{tprimedef}
T=a,
\end{equation}
since $a$ and $a^2$ only differ by a permutation of rows and columns and 
in the basis of Eq. (\ref{explicitd96}), it can be seen that multiplication by $c^\gamma d^\delta$ only yields phases 
which may be absorbed into the charged lepton fields.

Notice that the $T$ of Eq. (\ref{tprimedef}) can be diagonalised by the matrix,
\begin{equation}\label{Ve}
V^e=\frac{1}{\sqrt{3}}\begin{pmatrix}\omega^2&\omega&1\\\omega&\omega^2&1\\ 1&1&1\end{pmatrix},
\end{equation}
where $\omega=e^{2\pi i/3}$.
The ordering of the columns and rows in the above $V^e$ determines the ordering of the eigenvalues in $T$:
\begin{equation}\label{ttransform}
T \rightarrow V^{e\dagger}a V^e=\begin{pmatrix}\omega^2& 0&0 \\ 0&\omega&0 \\0 & 0&1\end{pmatrix}.
\end{equation}
For example, changing the order of the eigenvalues of $T$ by applying $a^\alpha$ to $T$ by $a^{\alpha\dagger}Ta^{\alpha}$ ($\alpha=1,2$) changes $V^e$ to $a^{\alpha} V^e $ which just permutes the rows of $V$ in Eq. (\ref{VPMNS}).  

Note that it is not always the case that the generators $S,T,U$ above generate the full $\Delta(6n^2)$ group.
It turns out that the Klein subgroup in Eq. (\ref{numklein2}), in combination with the residual $Z_3^T$ in Eq. (\ref{tprimedef}), will only generate the full $\Delta(6n^2)$ symmetry group if and only  if $\gamma'$ does not divide $n$.
From our top down point of view this is acceptable since we are only interested in the possible predictions that can arise from $\Delta(6n^2)$.

%%%%%%%%%%%%%%%%%%%%%%%%%%%%%%%%%%%%%%%%%%%%%%%%%%%%

\section{Results}

%%%%%%%%%%%%%%%%%%%%%%%%%%%%%%%%%%%%%%%%%%%%%%%%%%%%

Using the results of the previous section one can compute the columns of the lepton mixing matrix which correspond to each possible Klein subgroup of a certain $\Delta(6n^2)$ group where $n$ is even and we assume $T=a$. The steps for this procedure are summarised as follows. 

We shall generate all Klein group elements in  Eqs. (\ref{numklein1})-(\ref{numklein4})
in the explicit $\bf{3}$$_2^1$ representation matrices given in Eq. (\ref{explicitd96}), then transform each Klein group's elements to the basis where $T$ is diagonal via $V^e$, cf. Eq.(\ref{ttransform}). Here, the eigenvectors with +1 eigenvalue correspond to the columns of possible mixing matrices
as in Eq. (\ref{columns}). Since the ordering of the columns and rows of the mixing matrix calculated this way is arbitrary, without loss of generality we take the smallest absolute value from each mixing matrix and assign this as 
$V_{13}$ with its corresponding column being the third column eigenvector of $V$.   This completed procedure is unique up to interchanging the second and third rows of $V$, corresponding to two predictions for the atmospheric angle.

Implementing the preceding procedure for calculating the mixing matrix resulting from the Klein group in Eq. (\ref{numklein1}) with $T=a$ yields the \textit{old} trimaximal mixing matrix\cite{originaltrimax} which is given by the $V^e$ in Eq. (\ref{Ve}) up to permutation of its rows and columns. Clearly, this is not a phenomenologically viable mixing
matrix, so we discard this possibility.

We do not have to consider all the Klein groups in 
Eqs. (\ref{numklein2})-(\ref{numklein4}) since they all 
result in identical PMNS matrices up to permutations of rows and columns. This is because the Klein group elements in Eq. (\ref{numklein3}) and Eq. (\ref{numklein4}) are related to $G_i$ in Eq. (\ref{numklein2}) by 
$a^2 G_i a$ and $a G_ia^2$ respectively, where $a$ and  $a^2$ 
from Eq. (\ref{explicitd96}) interchange rows and columns.

Thus, it is sufficient to consider the Klein subgroup given in Eq. (\ref{numklein2}), where 
the element $c^{n/2}$ becomes the ``traditional'' $S$ generator in the basis in which $T$ is diagonal,
\begin{equation}
S \rightarrow V^{e\dagger}c^{n/2} V^e=\frac{1}{3}\begin{pmatrix}-1&2&2\\2&-1&2\\2&2&-1\end{pmatrix}
\end{equation}
This predicts one trimaximal middle column (TM$_2$), i.e. $(1,1,1)^T/\sqrt{3}$ \cite{TMnew}, in lepton mixing
\footnote{Note that a Klein symmetry corresponding to 
$V$ with a fixed column of $1/\sqrt{6}(2,-1,-1)^T$ (TM$_1$ mixing) 
cannot be identified as a subgroup of $\Delta(6n^2)$.}. 
This was also assumed in \cite{Holthausen:2012wt}.
The other elements of the same Klein subgroup also provide columns of $V$ which is then up to the order of rows and columns given by 
%\begin{equation}
%V=\begin{pmatrix}\frac{1}{\sqrt{6}N_+}(c+\sqrt{3} s+1)&\frac{1}{\sqrt{3}}&\frac{1}{\sqrt{6}N_-}(c+\sqrt{3} s-1)\\ 
%-\frac{N_+}{\sqrt{6}}&\frac{1}{\sqrt{3}}&\frac{N_-}{\sqrt{6}}\\
%\frac{1}{\sqrt{6}N_+}(2 c-1)&-\frac{1}{\sqrt{3}}&\frac{1}{\sqrt{6}N_-}(2 c+1)\end{pmatrix},
%\label{V}
%\end{equation}
\begin{eqnarray}
V=\left(
\begin{array}{ccc}
 \sqrt{\frac{2}{3}} \cos (\vartheta ) & \frac{1}{\sqrt{3}} & \sqrt{\frac{2}{3}} \sin (\vartheta ) \\
 -\sqrt{\frac{2}{3}} \sin \left(\frac{\pi }{6}+\vartheta\right) & \frac{1}{\sqrt{3}} & \sqrt{\frac{2}{3}} \cos \left(\frac{\pi
   }{6}+\vartheta\right) \\
 \sqrt{\frac{2}{3}} \sin \left(\frac{\pi }{6}-\vartheta \right) & -\frac{1}{\sqrt{3}} & \sqrt{\frac{2}{3}} \cos \left(\frac{\pi }{6}-\vartheta
   \right) \\
\end{array}
\right),
\label{V}
\end{eqnarray}
%where $c=\cos \vartheta$ and $s=\sin \vartheta$, with $\vartheta = 2 \pi \gamma^\prime/n$ and
%\begin{equation}
%N_\pm=\sqrt{2\pm \sqrt{3}s\mp c}.
%\end{equation} 
where $\vartheta=\pi\gamma'/n$  (cf.~\cite{Holthausen:2012wt}).  Since $\gamma' =1,\ldots, n/2$, we obtain discrete predictions for the mixing angles corresponding to 
$\vartheta =\pi /n, \ldots, \pi/2$.
In general we cannot predict the order of the rows and columns with this method, 
so we pick the entry with the smallest absolute value and assign it to be $|V_{13}|$.
Notice that for the different values of $\vartheta$,
%$\vartheta$
different elements of Eq. (\ref{V}) play the role of $V_{13}$. After $V_{13}$ has been fixed, the second and third row can still be interchanged, leading to two different predictions for the atmospheric angle, corresponding to 
$\delta_{CP}=0$ and $\delta_{CP}=\pi$, leading to the testable sum rules,
$\theta_{23}=45^\circ \mp \theta_{13}/\sqrt{2}$, respectively \cite{King:2013eh}.
(Note that Klein subgroups do not predict Majorana phases.)
These sum rule relations follow from considering the atmospheric angle sum rule 
given in \cite{King:2007pr} for the cases $\delta_{CP} =0, \pi$.
The sum rule $\theta_{23}=45^\circ - \theta_{13}/\sqrt{2}$ 
was also proposed in \cite{Harrison:2005dj} in a different context.

\begin{figure}[h]
\center
\includegraphics[width=0.8\textwidth]{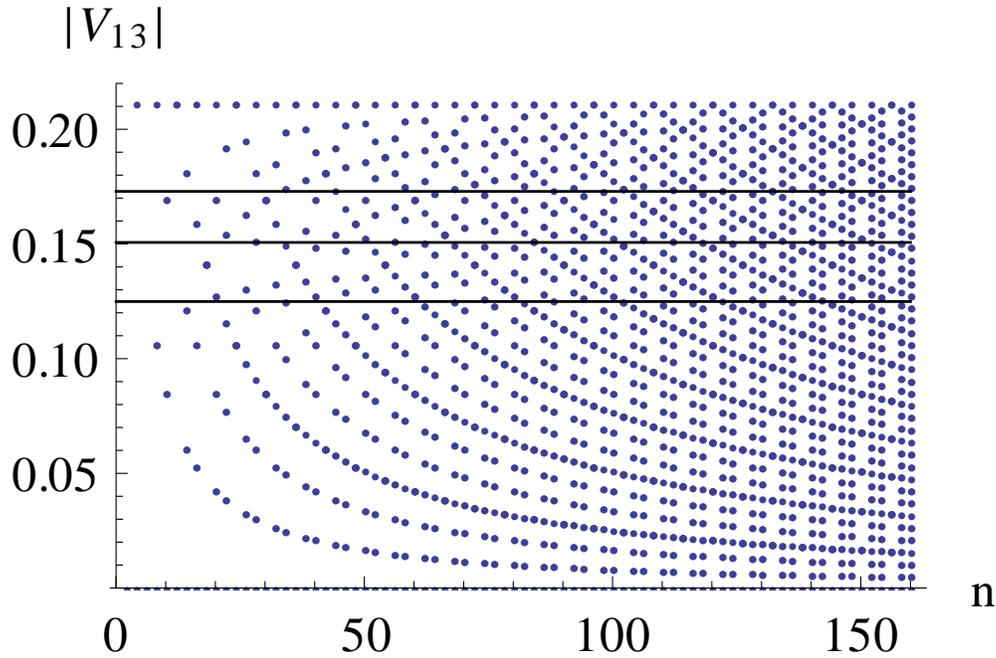}
\caption{\label{u13vsna} 
The possible values that $|V_{13}|$ can take in $\Delta(6n^2)$ family symmetry groups with even $n$. 
Examples include $|V_{13}|=0.211,0.170,0.160,0.154$ for $n=4,10,16,22$, respectively.
The lines denote the present approximate $3\sigma$ range of $|V_{13}|$.}
\end{figure}

FIG.~\ref{u13vsna} shows all possible predictions for $|V_{13}|$ corresponding to the different Klein subgroups for each $\Delta(6n^2)$ of even $n$ one obtains using the method previously discussed.
 As $n$ increases the number of possible values of $|V_{13}|$ predicted by $\Delta(6n^2)$ also increases
 according to the above discussion.

%%%%%%%%%%%%%%%%%%%%%%%%%%%%%%%%%%%%%%

\section{Conclusions}

%%%%%%%%%%%%%%%%%%%%%%%%%%%%%%%%%%%%%%%
In this paper we have obtained predictions of lepton mixing parameters for direct models based on 
$\Delta(6n^2)$ family symmetry groups for arbitrarily large $n$
in which the full Klein symmetry is identified as a subgroup
of the family symmetry.  After reviewing and developing the group theory associated with $\Delta(6n^2)$, we confirmed some known results of the recent numerical searches and found many new possible mixing patterns for large $n$
able to yield lepton mixing angle predictions within $3\sigma$ of recent global fits.
Previously, $\Delta(6n^2)$ had only been analysed
within particular scans up to a much lower order than we considered.
All the examples predict exact TM$_2$ mixing with oscillation phase zero or $\pi$ corresponding to two possible predictions for the atmospheric angle but differ in the prediction of $|V_{13}|$ as shown in FIG.~\ref{u13vsna}. 

For large $n$, it is clear that the predictions for $|V_{13}|$ densely fill the allowed
range. 
Nevertheless, our general method of analysing $\Delta(6n^2)$ family symmetry groups is of interest since 
it represents for the first time a model independent treatment of 
an infinite class of theories. The general predictions for the considered class of theories based on 
$\Delta(6n^2)$ are Majorana neutrinos,
trimaximal lepton mixing with reactor angle fixed up to a discrete choice, an oscillation phase of either zero or $\pi$ and sum rules 
$\theta_{23}=45^\circ \mp \theta_{13}/\sqrt{2}$, respectively, which are
consistent with the recent global fits and will be tested in the near future. 

\section{ACKNOWLEDGEMENTS}
The authors acknowledge partial support from the European Union FP7 ITN-INVISIBLES (Marie Curie Actions, PITN- GA-2011- 289442).  SFK and AJS acknowledge support from the STFC Consolidated ST/J000396/1 grant. SFK acknowledges support from EU ITN UNILHC PITN-GA-2009-237920.

\section{APPENDIX}
In this Appendix we show that $T$ generators of order greater than 3 are not viable.
We begin by considering the order of $T$ to be even. Then, $T^m=1$ with $m=2q$ where $q$ is an integer. 
We first note that diagonal $T$ candidates in the basis of Eq.~(\ref{explicitd96})
will not lead to acceptable mixing. After removing unphysical phases, all non-diagonal $T$ candidates 
of even order $m=2q$ can be written without loss of generality as,
\begin{equation}
T=bc^{\xi n/q}, \xi=1,\ldots, q-1 \label{evenT}.
\end{equation}
The matrices of Eq. (\ref{evenT}) are diagonalised by 
\begin{equation}
V^e=
\frac{1}{\sqrt{2}}\begin{pmatrix}
0 & e^{ -i \pi\xi/q } & -e^{- i \pi\xi/q}\\
 \sqrt{2} & 0 & 0 \\
 0 & 1 & 1
\end{pmatrix}.
\end{equation}
Applying the above matrix to $c^{n/2}$ results in:
\begin{equation}
U \rightarrow V^{e\dagger}c^{n/2} V^e=
\begin{pmatrix}
-1&0&0\\0&0&1\\0&1&0
\end{pmatrix}.
\end{equation}
The unique eigenvector of this generator is given by $(0,1,1)/\sqrt{2}$. Picking the smallest element of the mixing matrix as $V_{13}$ gives $V_{13}=0$. For $n=2$ this results in a completely 
bimaximal mixing matrix \cite{bimaximal}.  

If the order of $T$ is not even but can be divided by 3, application of a unitary transformation $R=c^xd^y$ can remove all phases implying only $T=a$ remains, yielding the previously discussed predictions for $T=a$.

Continuing the systematic consideration of candidate $T$ generators leads us to consider the case of a $T$ generator in which the order is odd, not divisible by 3 but larger than 3.  A $\Delta(6n^2)$ group can only contain such an element if $m$ divides $n$. Then, for this case the possible $T$ generators are given by 
\begin{equation}
T=c^{\mu n/m}d^{\rho n/m}
\end{equation}
where $\mu,\rho=0,\ldots, m-1$ and $\mu,\rho$ are not simultaneously zero.
These yield no phenomenologically viable predictions.  Therefore, only candidate $T$ generators from $Z_3$ subgroups of $\Delta(6n^2)$ are phenomenologically viable.

% \bibliography{Delta6nsquared_final2}

%%%%%%%%%%%%%%%%%%%%%%%%%%%%%%Trimaximal Mixing

\end{document}